\documentclass[useAMS,usegraphicx,usenatbib,letters]{mn2e}
\usepackage{lscape}

\title{A new cold sub-Saturnian candidate planet orbiting GJ 221}

\author[M. Tuomi 2013]{Mikko Tuomi\thanks{E-mail: \texttt{mikko.tuomi@utu.fi}; \texttt{m.tuomi@herts.ac.uk}}
\\
University of Hertfordshire, Centre for Astrophysics Research, Science and Technology Research Institute, College Lane, AL10 9AB,\\ Hatfield, UK\\
University of Turku, Tuorla Observatory, Department of Physics and Astronomy, V\"ais\"al\"antie 20, FI-21500, Piikki\"o, Finland}

\begin{document}

\date{Accepted {XX.XX.2014}. Received {XX.XX.2014}; in original form {XX.XX.2014}}

\pagerange{\pageref{firstpage} -- \pageref{lastpage}} \pubyear{2014}

\maketitle

\label{firstpage}

\begin{abstract}
We re-analyse the recently published HARPS and PFS velocities of the nearby K dwarf GJ 221 that have been reported to contain the signatures of two planets orbiting the star. Our goal is to see whether the earlier studies discussing the system fell victims of false negative detections. We perform the analyses by using an independent statistical method based on posterior samplings and model comparisons in the Bayesian framework that is known to be more sensitive to weak signals of low-mass planets. According to our analyses, we find strong evidence in favour of a third candidate planet in the system corresponding to a cold sub-Saturnian planet with an orbital period of 500 days and a minimum mass of 29 $M_{\oplus}$. Application of sub-optimal signal detection methods can leave low-amplitude signals undetected in radial velocity time-series. Our results suggest that the estimated statistical properties of low-mass planets can thus be biased because several signals corresponding to low-mass candidate planets may have gone unnoticed. This also suggests that the occurrence rates of such planets based on radial velocity surveys might be underestimated.
\end{abstract}


% ----------------------------------------------------------------

\section{Introduction}

Metal-poor low-mass M and K dwarfs are commonly hosts to diverse planetary systems populated by planets with masses below that of Saturn, whereas more massive planets are not found orbiting them in abundance \citep[e.g.][]{howard2010,howard2012,mayor2011,santos2011,bonfils2013a,tuomi2013_M}. Examples of such systems are e.g. the K dwarfs HD 40307, which hosts a six-planet system \citep{mayor2009,tuomi2013a}, and HD 69830 with three sub-Neptunian planets orbiting it \citep{lovis2006}. Both these systems have been detected by using the Doppler spectroscopy method whereas another such example, the Kepler-62 five-planet system, has been detected by the transit photometry method \citep{borucki2013}. Moreover, M dwarfs are also known to host diverse systems of low-mass planets \citep[e.g.][]{mayor2009b,anglada2013,bonfils2013b,tuomi2013c} and such systems appear to be very common in the Solar neighbourhood \citep{dressing2013,bonfils2013a,tuomi2013_M}.

Another similar target is the metal-depleted ([Fe/H] $= -0.07 \pm 0.10$) nearby K7 or M0 dwarf GJ 221 with two known planets orbiting it \citep{locurto2013,arriagada2013}. The first to report this detection, made by using the HARPS spectrograph and its standard cross-correlation function (HARPS-CCF) spectral reduction applied to 102 spectra, was \citet{locurto2013}. This detection was soon verified by taking advantage of the HARPS-TERRA algorithm that provides more accurate velocities for M and K dwarfs \citep{anglada2012} that enabled the detection of the two planet candidates by using the HARPS velocities of only 61 spectra and 26 somewhat less precise radial velocities from PFS \citep{arriagada2013}. Although the 61 spectra analysed by \citet{arriagada2013} were included in the analyses of \citet{locurto2013}, the two studies are independent with respect to the different spectral reduction methods used.

The studies of \citet{locurto2013} and \citet{arriagada2013} relied on identifying powers in the periodograms of the velocity time-series exceeding the detection thresholds defined by analytical or numerical false alarm probabilities. However, such analyses of unevenly spaced time-series that contain excess noise that is not distributed according to the Gaussian density and is not white are prone to false positives \citep{vogt2010,tuomi2011} and negatives \citep{tuomi2013c}.\footnote{See also the discussions in \citet{tuomi2012} and \citet{tuomi2013b} underlining the shortcomings of periodogram analyses.} Instead, statistical models taking into account correlated noise \citep[e.g.][]{baluev2009,baluev2012,tuomi2013b,tuomi2013_M} and searches of the parameter space by using some well-established posterior sampling algorithms \citep[e.g.][]{metropolis1953,hastings1970,haario2001} provide a more robust approach for the detection of signals that do not show as significant periodogram powers. Moreover, a smart definition of prior densities of model parameters \citep[see e.g.][]{kipping2013,tuomi2013c} can be used very efficiently in decreasing the probability of Keplerian signals with high eccentricities that cannot be expected to correspond to signals of planetary origin in systems of multiple planets \citep{anglada2013,tuomi2013c}. Moreover, the combination of TERRA velocities and such techniques enabled \citet{tuomi2013c} to detect the same three planet candidates orbiting GJ 163 with roughly $\sim$ 35\% of the data that \citet{bonfils2013b} had in their independent analysis. \citet{locurto2013} and \citet{arriagada2013} also searched for signals in data residuals that were calculated based on the assumption that no additional signals exist in the data, thus biasing the results towards a lower number of signals \citep[see also][]{anglada2013,feroz2013}. This suggests that the currently accumulating high-precision spectra and the corresponding radial velocity data sets might contain signals of additional currently unknown planet candidates that have simply gone unnoticed due to applications of sub-optimal signal detection techniques.

In the current work, we present our re-analysis of the HARPS-CCF velocity data of GJ 221 \citep{locurto2013} in combination with the PFS data of \citet{arriagada2013}. Moreover, we present the corresponding re-analysis of the HARPS-TERRA and PFS velocities of \citet{arriagada2013} to verify the results obtained by the HARPS-CCF data. We search for additional signals in these data sets indicative of additional planet candidates orbiting the star. These analyses are performed following \citet{tuomi2013c} and by applying the signal detection criteria of \citet{tuomi2012} that have been used in several studies such as \citet{anglada2012b}, \citet{anglada2013}, \citet{tuomi2013a}, and \citet{tuomi2013b}. Thus, for a positive detection of a signal, we require that (1) a model containing this signal is at least 150 times more probable than a model without it, that (2) the period of this signal is well-constrained from above and below such that it can be called a periodic signal, and that (3) the amplitude of this signal is statistically significantly different from zero.

\section{Results}

\begin{table*}
\caption{Three-Keplerian solution of the combined HARPS-CCF and PFS velocities (top) and the combined HARPS-TERRA and PFS velocities (bottom). MAP estimates and the 99\% BCSs.}\label{tab:parameters}
\begin{center}
\begin{tabular}{lccc}
\hline \hline
Parameter & GJ 221 b & GJ 221 c & GJ 221 d \\
\hline
$P$ [days] & 3.8731 [3.8723, 3.8739] & 125.4 [124.8, 126.0] & 496 [474, 525] \\
$K$ [ms$^{-1}$] & 4.06 [2.94, 5.31] & 8.36 [6.54, 10.37] & 3.17 [1.54, 4.97] \\
$e$ & 0.03 [0, 0.20] & 0.16 [0, 0.32] & 0.03 [0, 0.25] \\
$\omega$ [rad] & 0.3 [0, $2\pi$] & 1.3 [0.0, 2.8] & 3.5 [0, $2\pi$] \\
$M_{0}$ [rad] & 6.0 [0, $2\pi$] & 0.4 [4.9, 2.0] & 4.1 [0, $2\pi$] \\
$m_{p} \sin i$ [M$_{\oplus}$] & 7.3 [5.1, 9.9] & 49 [36, 62] & 30 [14, 47] \\
$a$ [AU] & 0.0415 [0.0394, 0.0435] & 0.423 [0.400, 0.442] & 1.06 [0.99, 1.13] \\
\hline
$P$ [days] & 3.8730 [3.8721, 3.8740] & 125.2 [124.5, 125.7] & 500 [474, 532] \\
$K$ [ms$^{-1}$] & 3.46 [2.17, 4.62] & 8.30 [6.57, 9.86] & 3.10 [1.22, 4.97] \\
$e$ & 0.03 [0, 0.27] & 0.10 [0, 0.25] & 0.04 [0, 0.29] \\
$\omega$ [rad] & 1.6 [0, $2\pi$] & 2.6 [0, $2\pi$] & 4.8 [0, $2\pi$] \\
$M_{0}$ [rad] & 1.1 [0, $2\pi$] & 1.6 [0, $2\pi$] & 5.6 [0, $2\pi$] \\
$m_{p} \sin i$ [M$_{\oplus}$] & 6.3 [3.8, 8.6] & 48 [37, 59] & 29 [9, 47] \\
$a$ [AU] & 0.0415 [0.0395, 0.0436] & 0.421 [0.401, 0.443] & 1.06 [0.99, 1.13] \\
\hline \hline
\end{tabular}
\end{center}
%\tablefoot{}
\end{table*}

Following \citet{tuomi2013b} and \citet{tuomi2013c} we analysed the data sets with a statistical model that accounts for Keplerian signals and a linear trend; zero-mean Gaussian white noise with a variance of $\sigma^{2} = \sigma_{i}^{2} + \sigma_{l}^{2}$, where $\sigma_{i}$ is the estimated instrument uncertainty and $\sigma_{l}$ represents the excess white noise in the $l$th instrument; and correlations between the noise of the $i$th and $i+1$th measurement with a coefficient of $\phi \exp \big[ (t_{i} - t_{i+1})/\tau \big]$, where the exponential smoothing time-scale was set to $\tau = 4$ days. We chose this time-scale based on results of \citet{tuomi2013_M} because treating it as a free parameter did not affect the results significantly. We use the priors as described in \citet{tuomi2013c}, Section 5.4., i.e. an eccentricity prior that is proportional to a Gaussian function with a zero mean and a variance of $0.1^{2}$ and flat priors for the other parameters. We note that we use $l = \log P$ as a parameter in the samplings instead of the period $P$ as such. Therefore, the prior of the period is uniform in $\log P$.

\subsection{Combination of HARPS-CCF and PFS data}

\begin{figure}
\center
\includegraphics[angle=-90,width=0.45\textwidth]{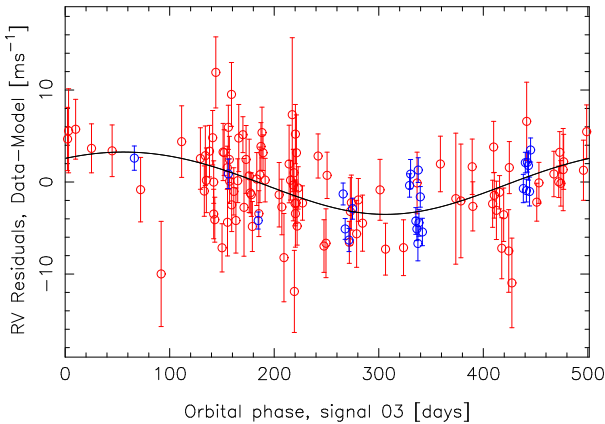}
\caption{Phase-folded signal of GJ 221 d given the combined HARPS-CCF (red circles) and PFS (blue) data sets.}\label{fig:CCF_signal}
\end{figure}

The HARPS-CCF data published by \citet{locurto2013} consists of 102 velocity measurements with a baseline of 2955 days. We obtained three periodic signals from the combined HARPS-CCF and PFS velocities at periods of 3.87, 125, and 500 days with velocity amplitudes of 4.1, 8.8, and 3.0 ms$^{-1}$, respectively. The maximum \emph{a posteriori} (MAP) parameter estimates and the corresponding 99\% Bayesian credibility intervals (or Bayesian credibility sets, BCSs) are shown in Table \ref{tab:parameters}. The three signals are detected very confidently and satisfy the detection criteria of \citet{tuomi2012}. This can be seen by looking at the 99\% BCSs of the orbital periods and velocity amplitudes in Table \ref{tab:parameters} that are clearly well-constrained from above and below. This can also be seen in Fig. \ref{fig:CCF_signal}, where we have plotted the phase-folded signal. We also demonstrate that the signals are very significant by showing the log-Bayesian evidences and the derived log-Bayesian evidence ratios in Table \ref{tab:probabilities}. According to these evidence ratios, the three-Keplerian model is roughly 2.6$\times 10^{4}$ times more probable (when assuming equal \emph{a priori} model probabilities) than the two-Keplerian model, which exceeds even the conservative detection threshold of $10^{4}$ applied by \citet{tuomi2013_M}. We note that this threshold corresponds to an evidence ratio of 9.2.

\begin{table}
\caption{Log-Bayesian evidence ratios ($\ln B_{i,j} = \ln P(m | \mathcal{M}_{j}) - \ln P(m | \mathcal{M}_{i})$) for models with $k = 0, ..., 3$ Keplerian signals given the combined PFS and HARPS-CCF data of \citet{locurto2013} ($m_{L}$) and the combined PFS and HARPS-TERRA data of \citet{arriagada2013} ($m_{A}$).}\label{tab:probabilities}
\begin{center}
\begin{tabular}{lcc}
\hline \hline
  & $m_{L}$ & $m_{A}$\\
\hline
 $\ln B_{0,1}$ & 38.0 & 38.1 \\
 $\ln B_{1,2}$ & 28.4 & 16.0 \\
 $\ln B_{2,3}$ & 10.1 & 10.5 \\ 
 \hline \hline
\end{tabular}
\end{center}
%\tablefoot{}
\end{table}

\begin{figure}
\center
\includegraphics[angle=270, width=0.23\textwidth]{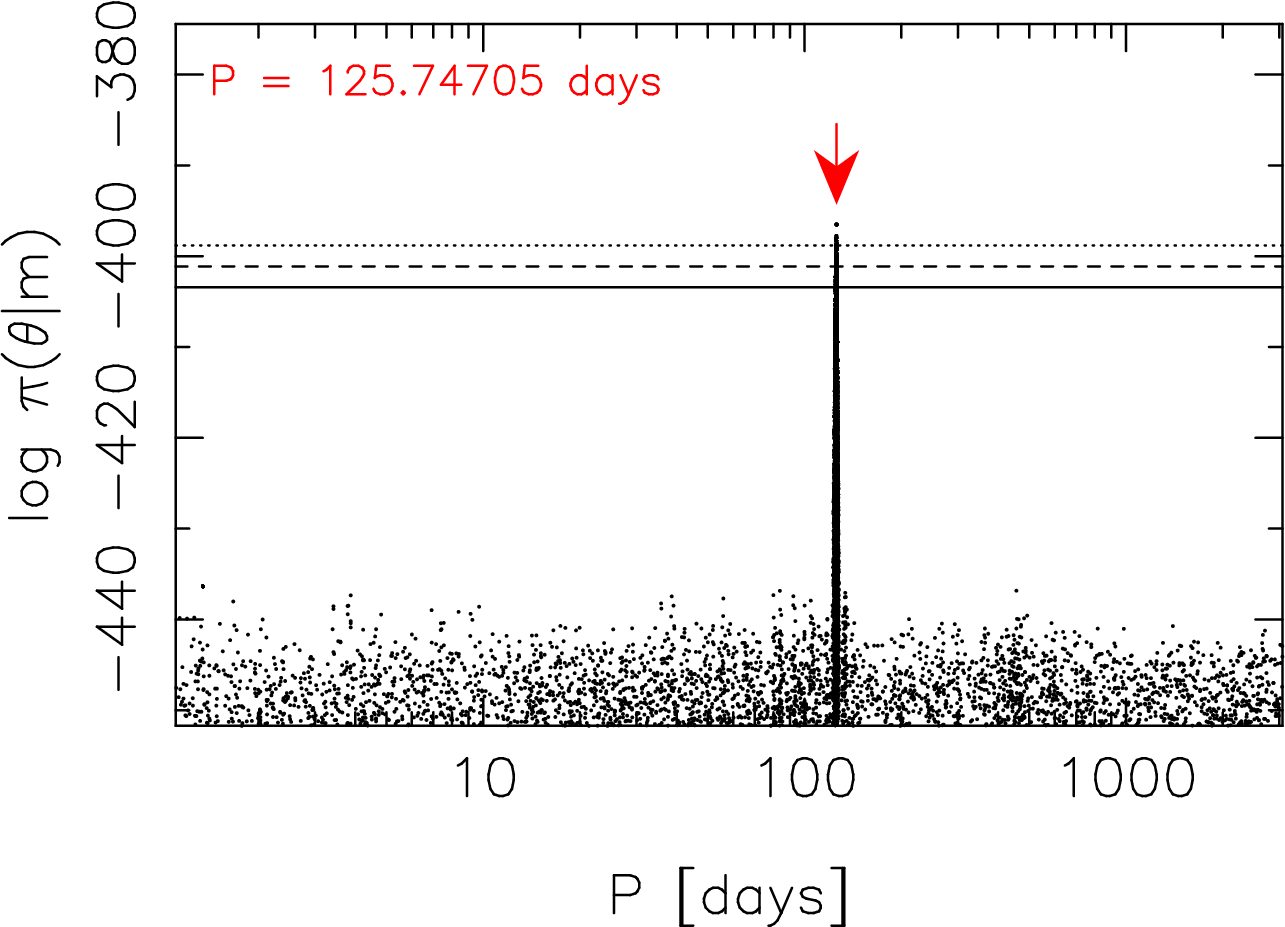}
\includegraphics[angle=270, width=0.23\textwidth]{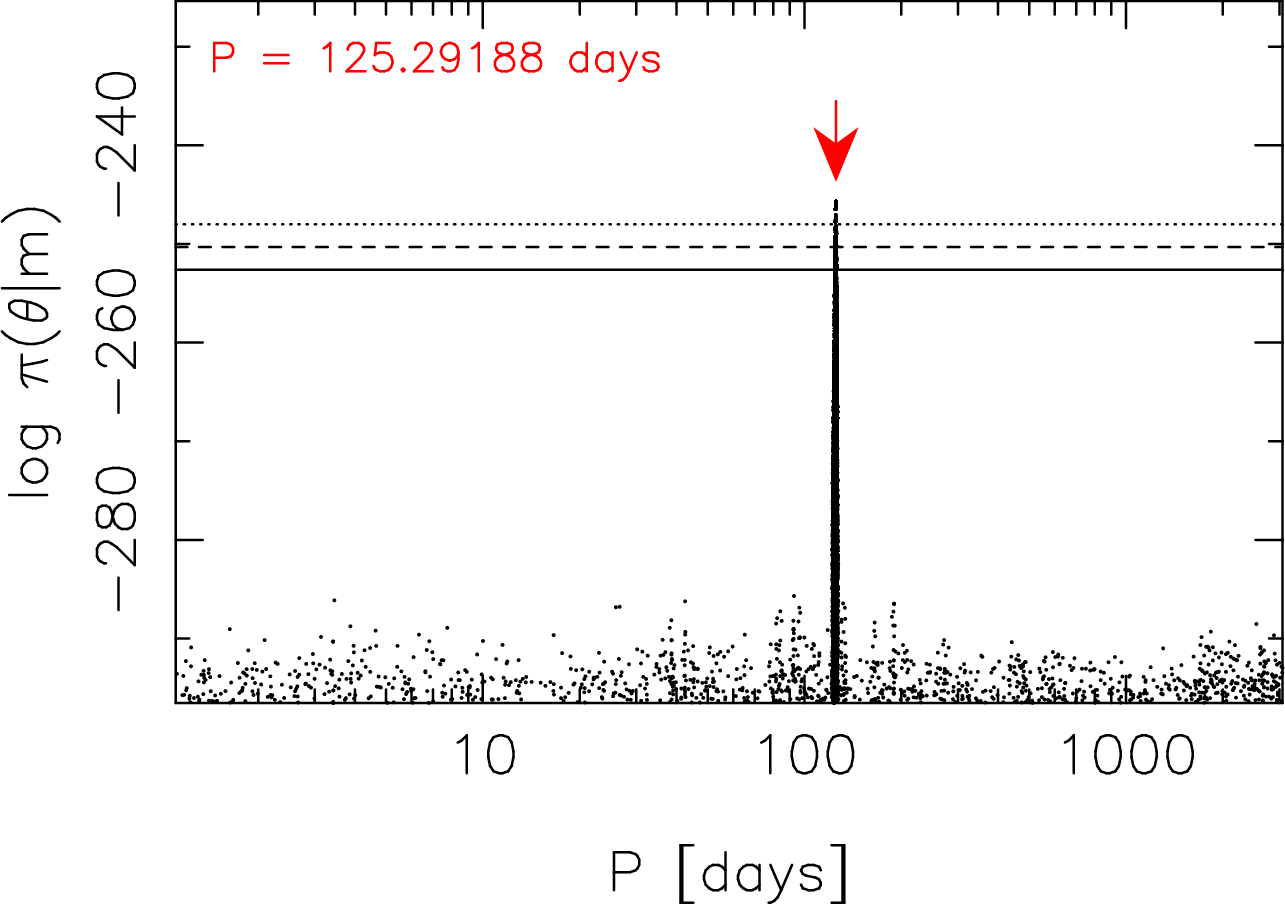}

\includegraphics[angle=270, width=0.23\textwidth]{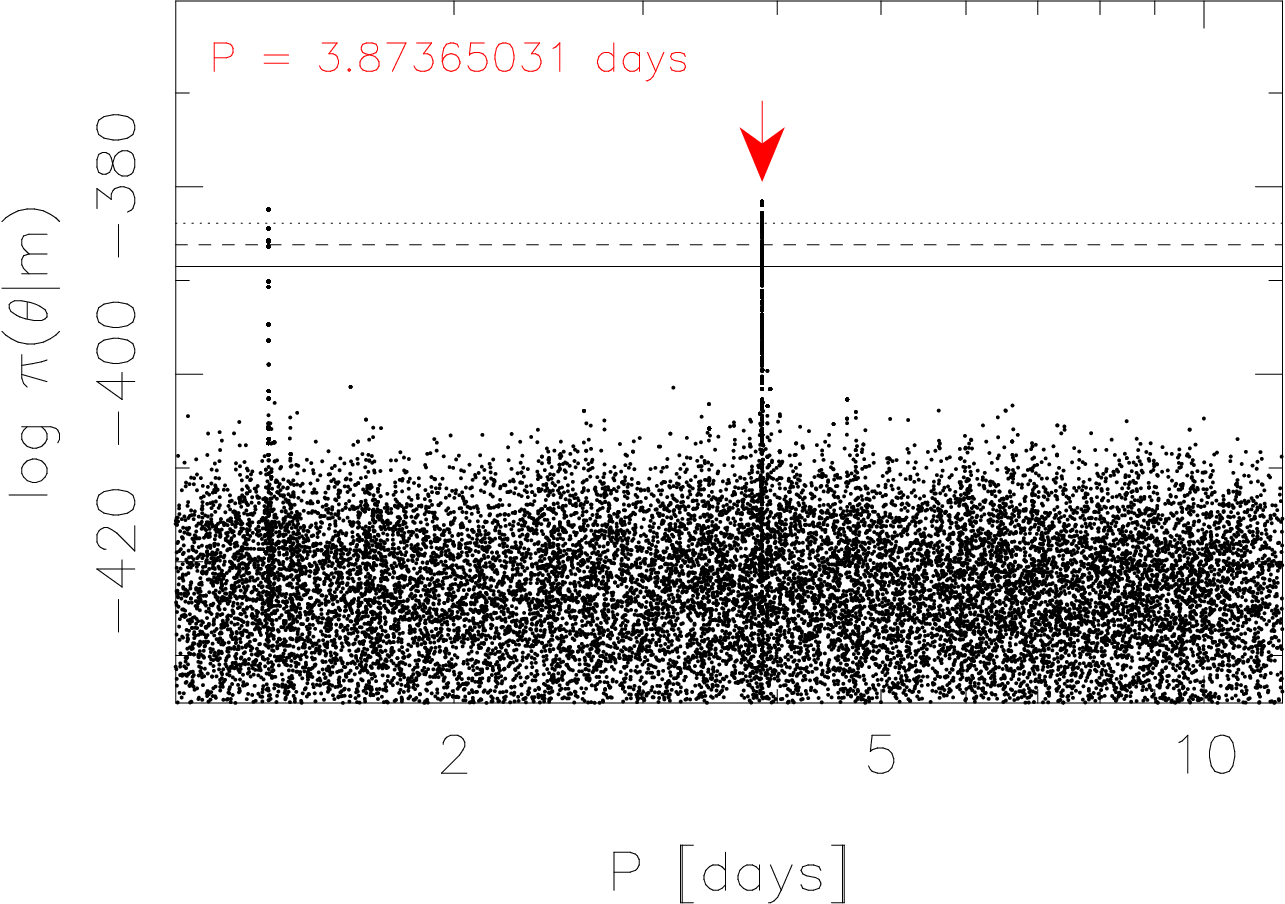}
\includegraphics[angle=270, width=0.23\textwidth]{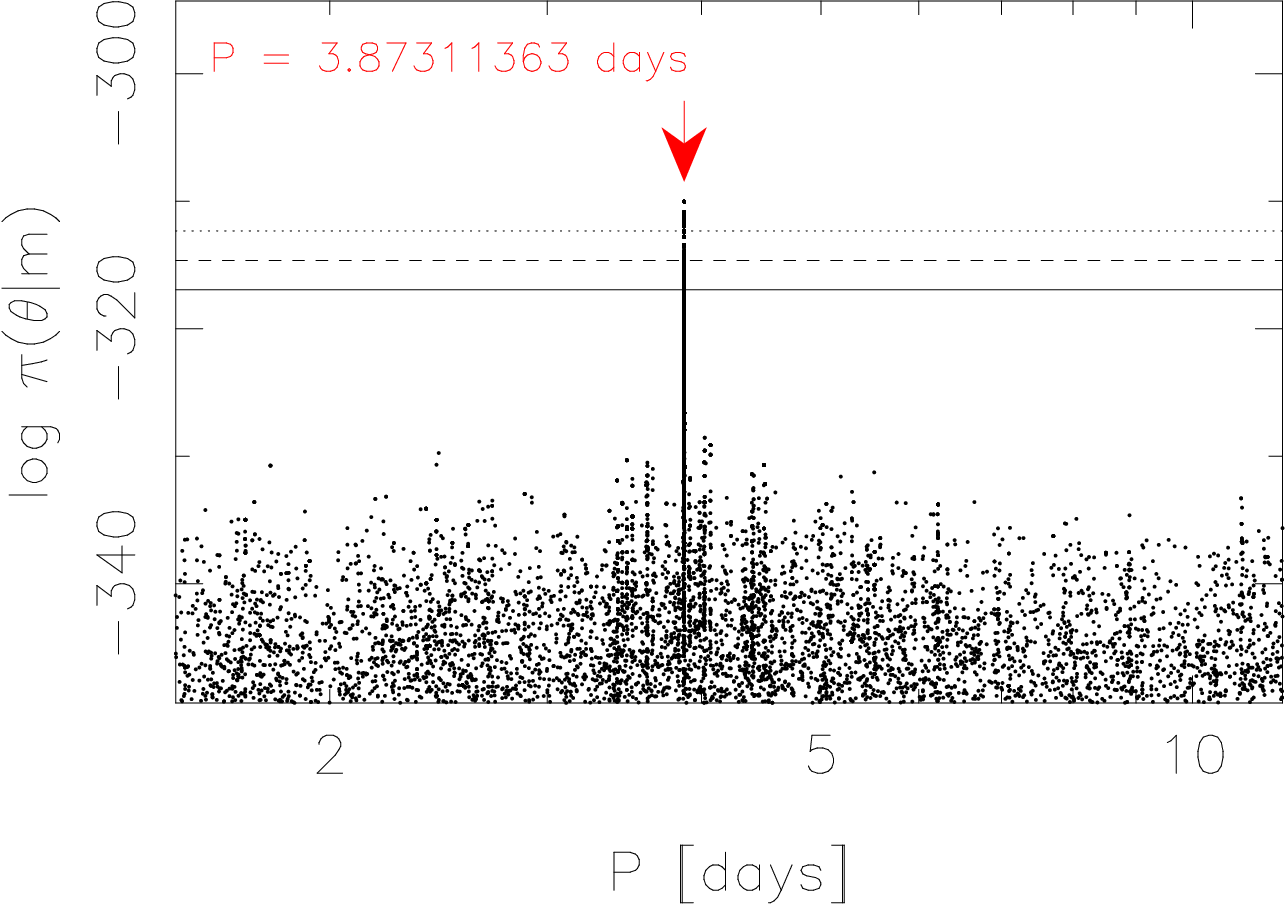}

\includegraphics[angle=270, width=0.23\textwidth]{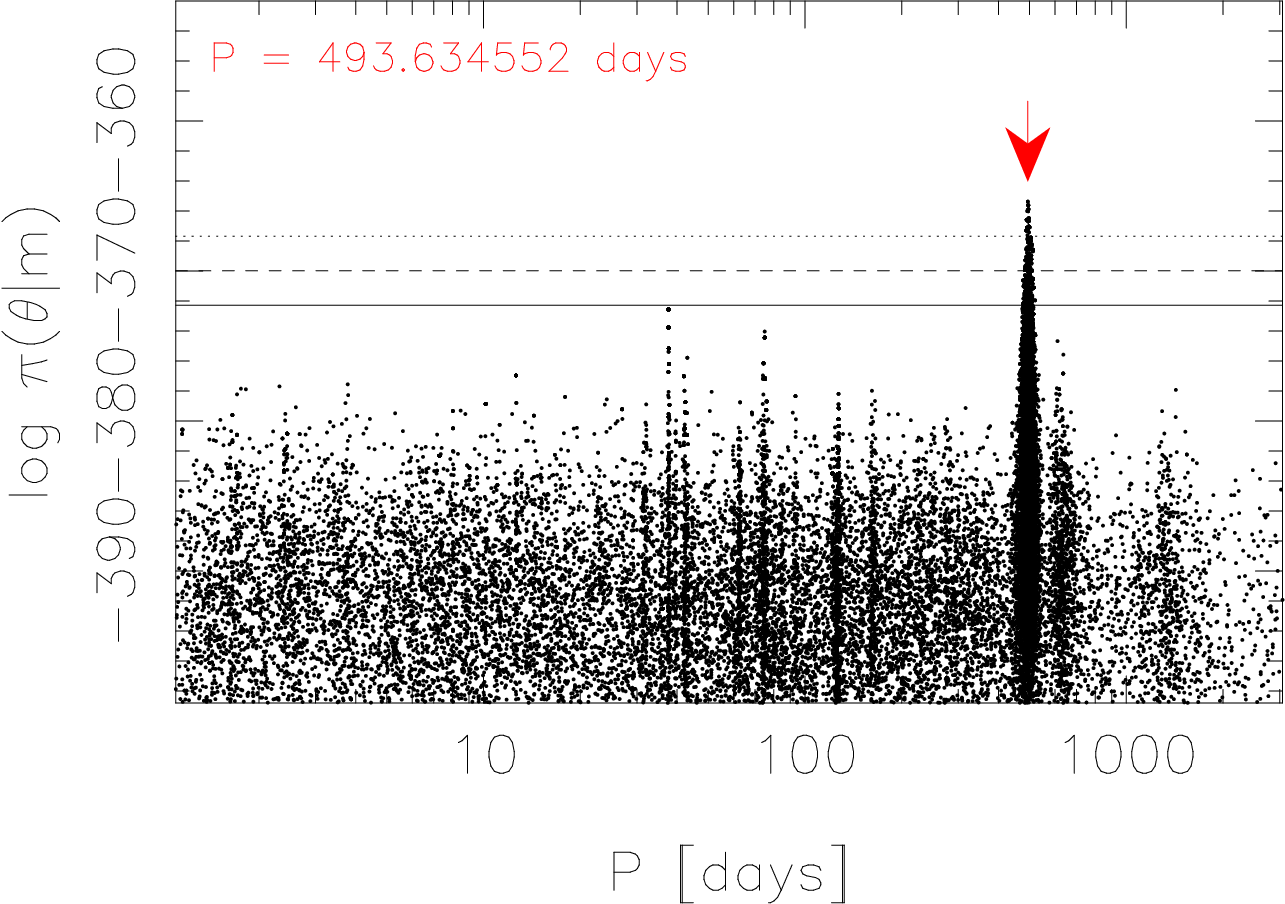}
\includegraphics[angle=270, width=0.23\textwidth]{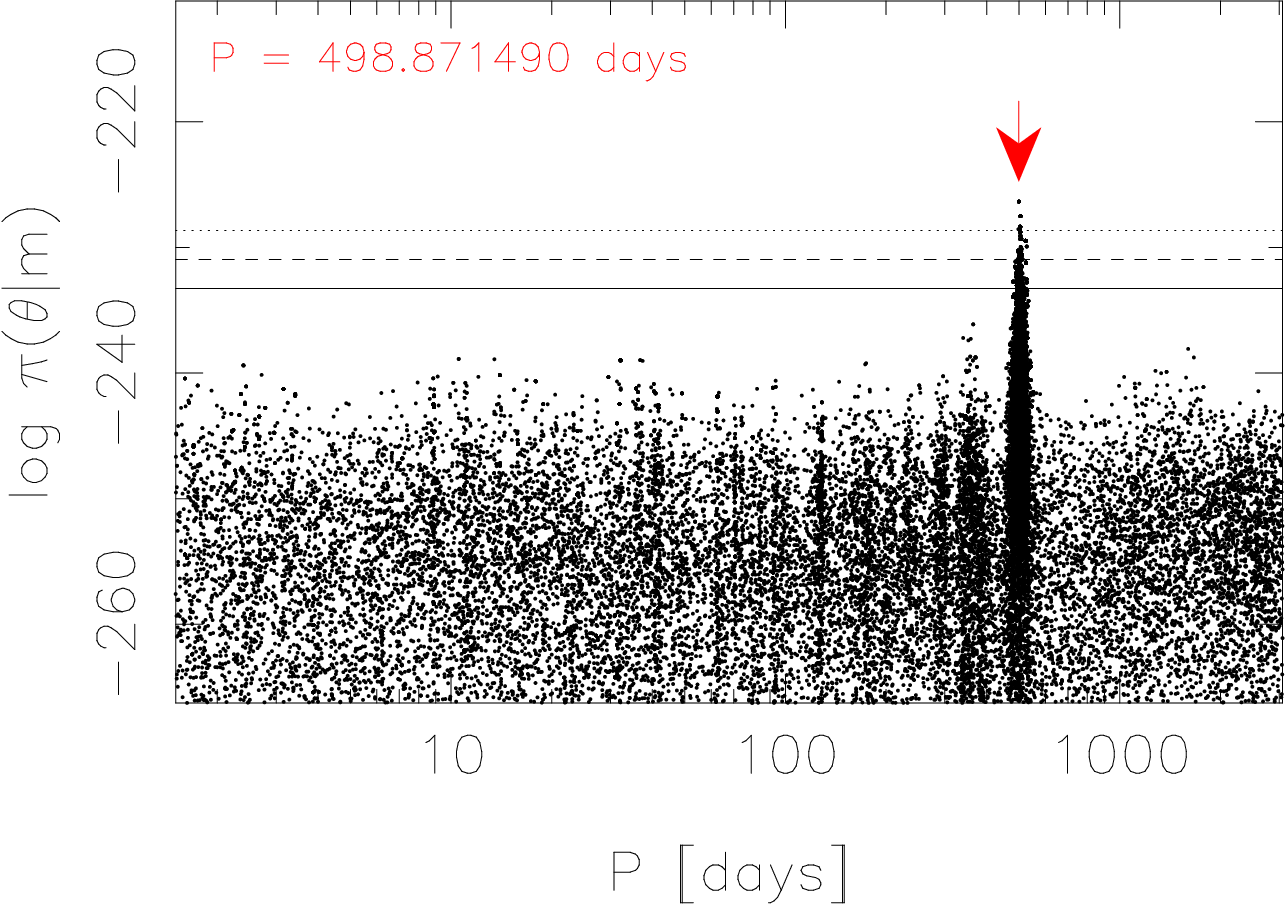}
\caption{Estimated log-posterior densities from posterior samplings as functions of the periods of the three Keplerian signals: GJ 221 b (top), GJ 221 c (middle), and GJ 221 d (bottom). Left panels indicate the densities given the combined HARPS-CCF and PFS data and right panels given the combined HARPS-TERRA and PFS data. The dotted, dashed, and solid lines denote the 10\%, 1\%, and 0.1\% probability thresholds of the maximum values in the samplins that are indicated by the red arrows. Note that these values are not necessarily equal to the MAP estimates because they are based on tempered samplings and therefore on scaled posterior densities.}\label{fig:log-posteriors}
\end{figure}

Searching the period space of the models with $k=1, 2,$ and 3 Keplerian signals is the key in identifying the periodic signals in the radial velocity data \citep[e.g.][]{tuomi2013c,tuomi2013_M}. We demonstrate such searches by showing some of the samples we have drawn from the posterior density with tempered posterior samplings, i.e. samplings where $\pi^{\beta}$ is used instead of the standard posterior density $\pi$ for some $\beta \in (0,1)$ \citep[see also][]{gregory2005}. We have plotted the resulting log-posterior densities as functions of period for each of the three signals we have detected in the combined HARPS-CCF and PFS data in Fig. \ref{fig:log-posteriors}. As can be seen, the 125-day signal stands out as a considerable maximum (Fig. \ref{fig:log-posteriors}, top left panel), whereas the other two signals are less convincingly, yet demonstrably, present as global maxima in the period space (middle and bottom left panels).

We also spotted correlation in the HARPS-CCF velocities \citep[see e.g.][]{tuomi2013b,tuomi2013c} with an estimated coefficient of $\phi =$ 0.56 [0.0, 1] that was not accounted for by \citet{locurto2013}. Furthermore, we observed a linear trend in the combined HARPS-CCF and PFS data of -1.05 [-1.70, -0.40] ms$^{-1}$year$^{-1}$ that \citet{locurto2013} do not discuss either and that could be an indication of a long-period sub-stellar companion orbiting the star or related to changing stellar activity \citep{dumusque2011,locurto2013}. The excess white noise ($\sigma_{l}$) in the HARPS-CCF velocities was found to have an MAP estimate of 1.96 ms$^{-1}$ and a 99\% BCS of [0.70, 3.23] ms$^{-1}$. The corresponding estimate for PFS data is 2.04 [0.93, 3.67] ms$^{-1}$. These estimates indicate that the signal of GJ 221 d with an amplitude of 3.17 [1.54, 4.97] ms$^{-1}$ should indeed be detectable in the data as it is likely higher than the radial velocity noise.

\subsection{Combination of HARPS-TERRA and PFS data}

The HARPS-TERRA data of \citet{arriagada2013} consist only of 61 velocity measurements and a baseline of 2150 days because all the spectra of \citet{locurto2013} were not available for \citet{arriagada2013} in the European Southern Observatory archive. Compared to the HARPS-CCF data, the number of measurements is roughly 60\% with a baseline of roughly 70\%. This suggests that the HARPS-TERRA data might not be sensitive to all the three signals detected from the HARPS-CCF data. However, this is not necessarily the case because the TERRA velocities are also known to be more precise than the CCF velocities for K and M dwarfs \citep{anglada2012,anglada2012b,tuomi2013c}.

\begin{figure*}
\center
\includegraphics[angle=-90,width=0.19\textwidth]{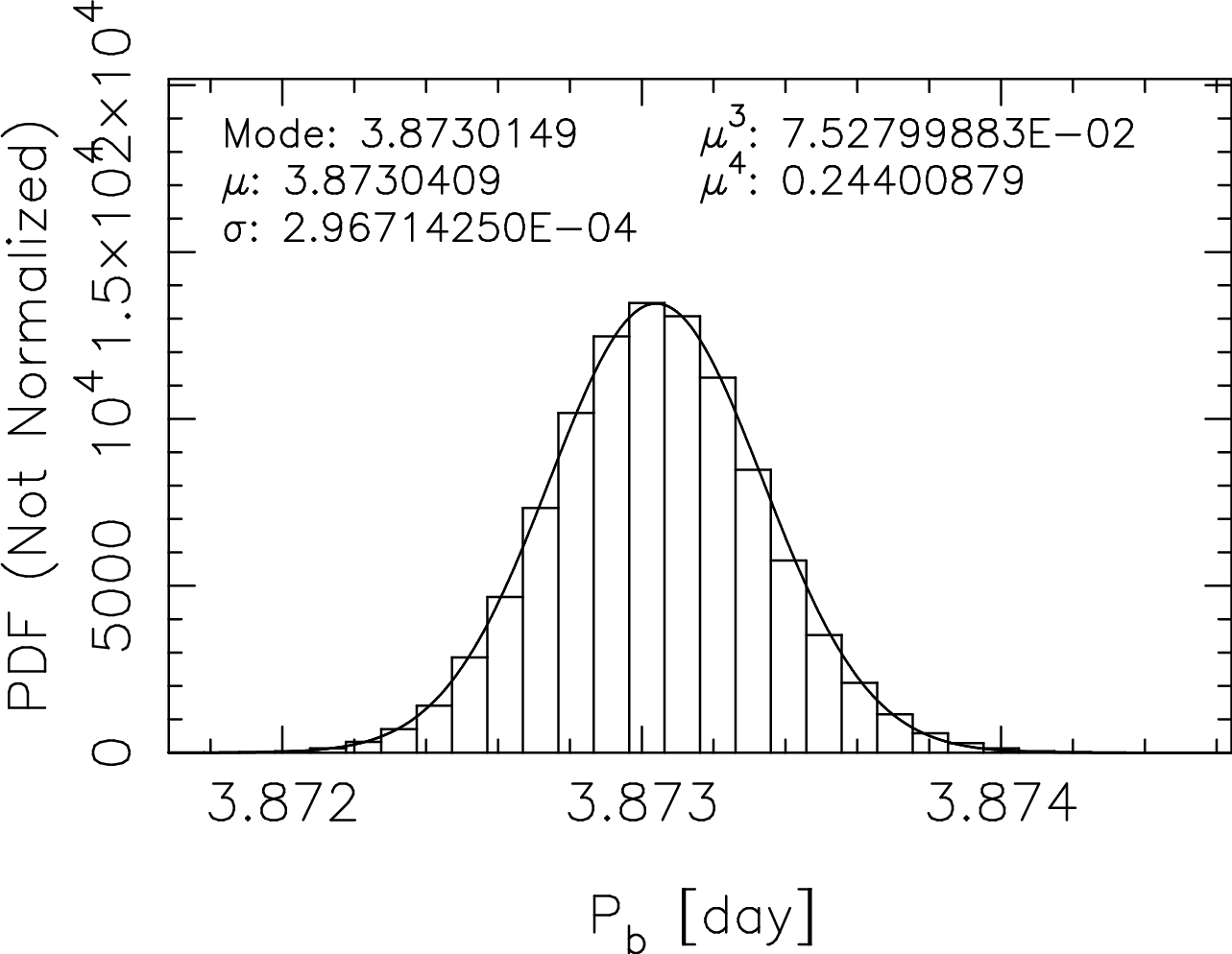}
\includegraphics[angle=-90,width=0.19\textwidth]{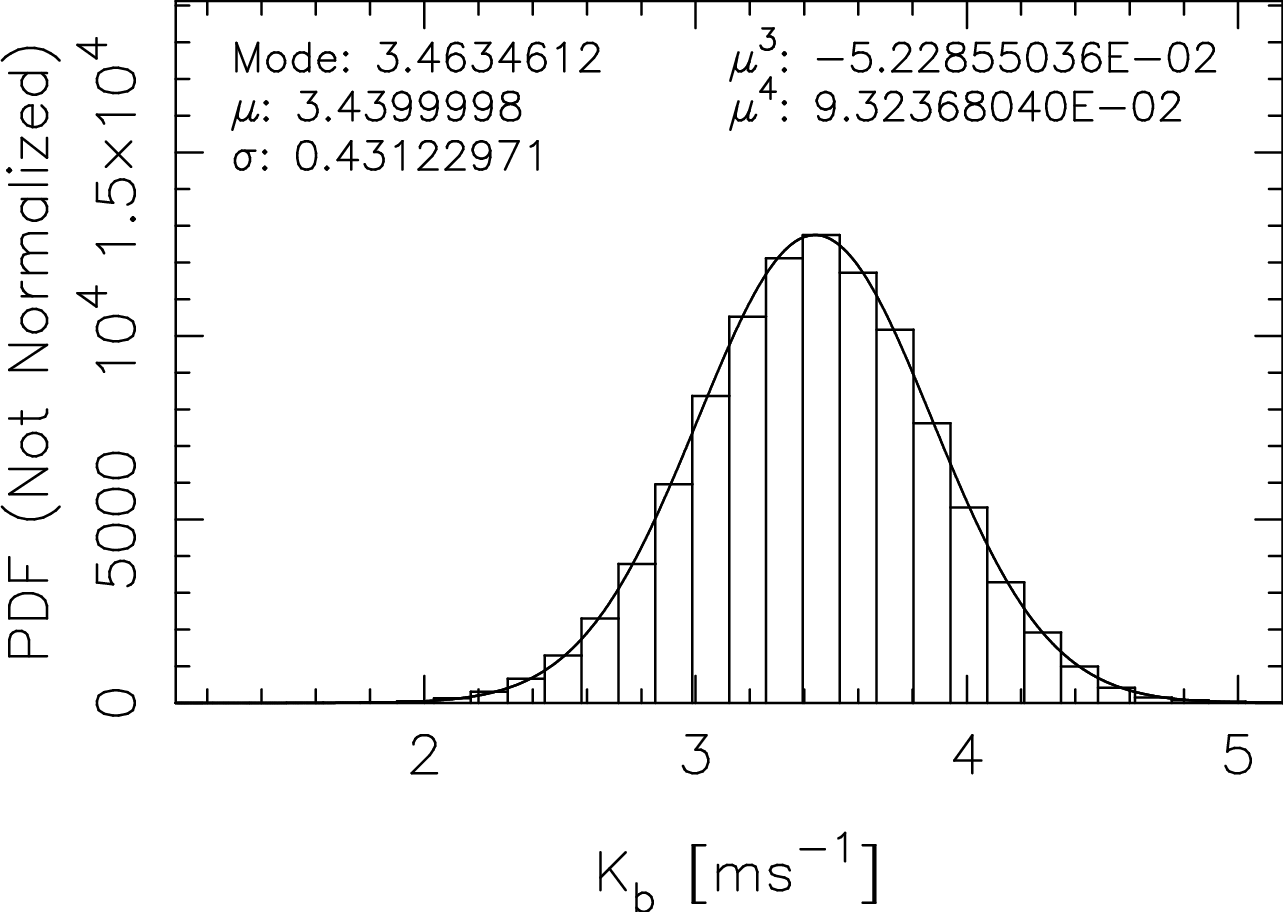}
\includegraphics[angle=-90,width=0.19\textwidth]{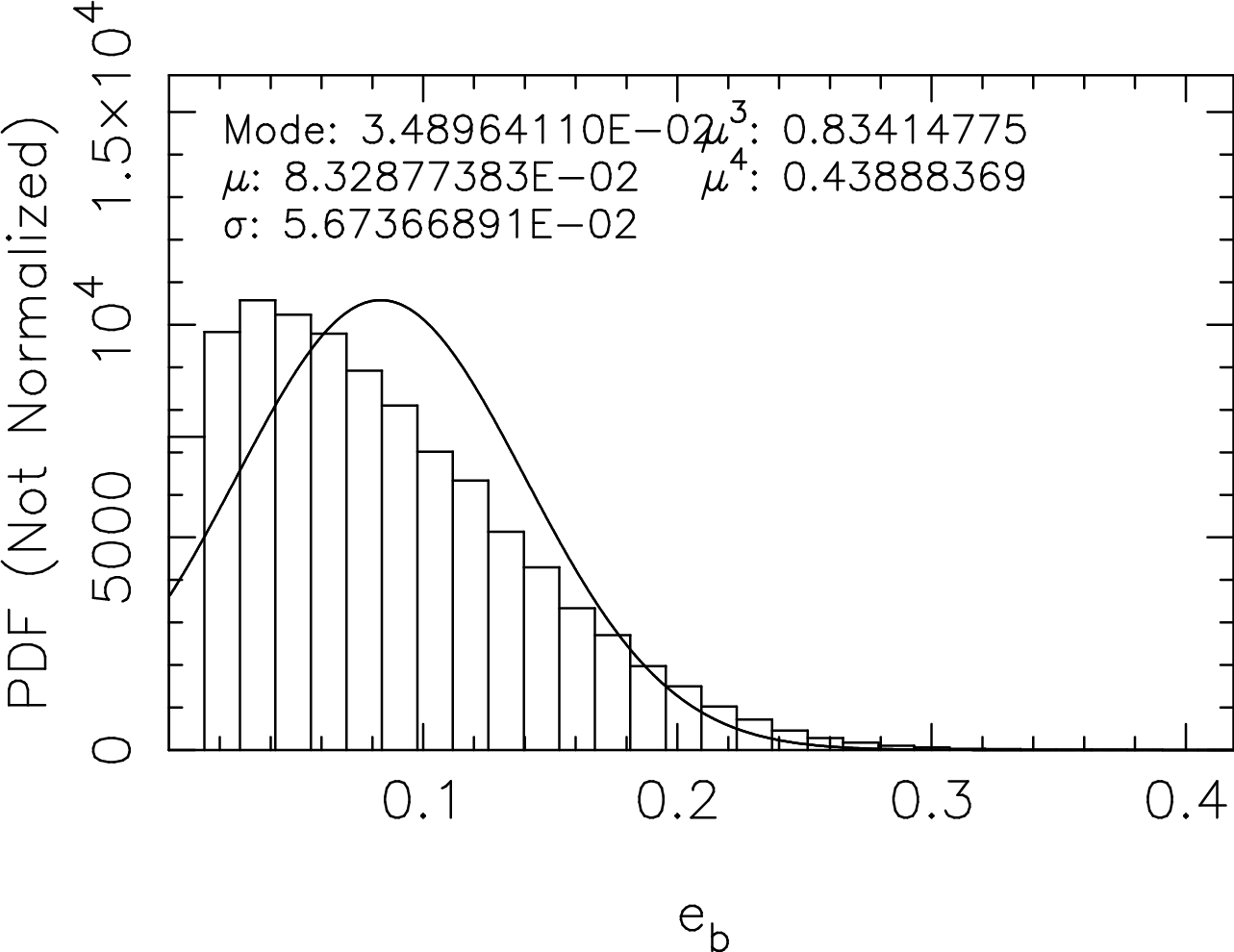}
\includegraphics[angle=-90,width=0.19\textwidth]{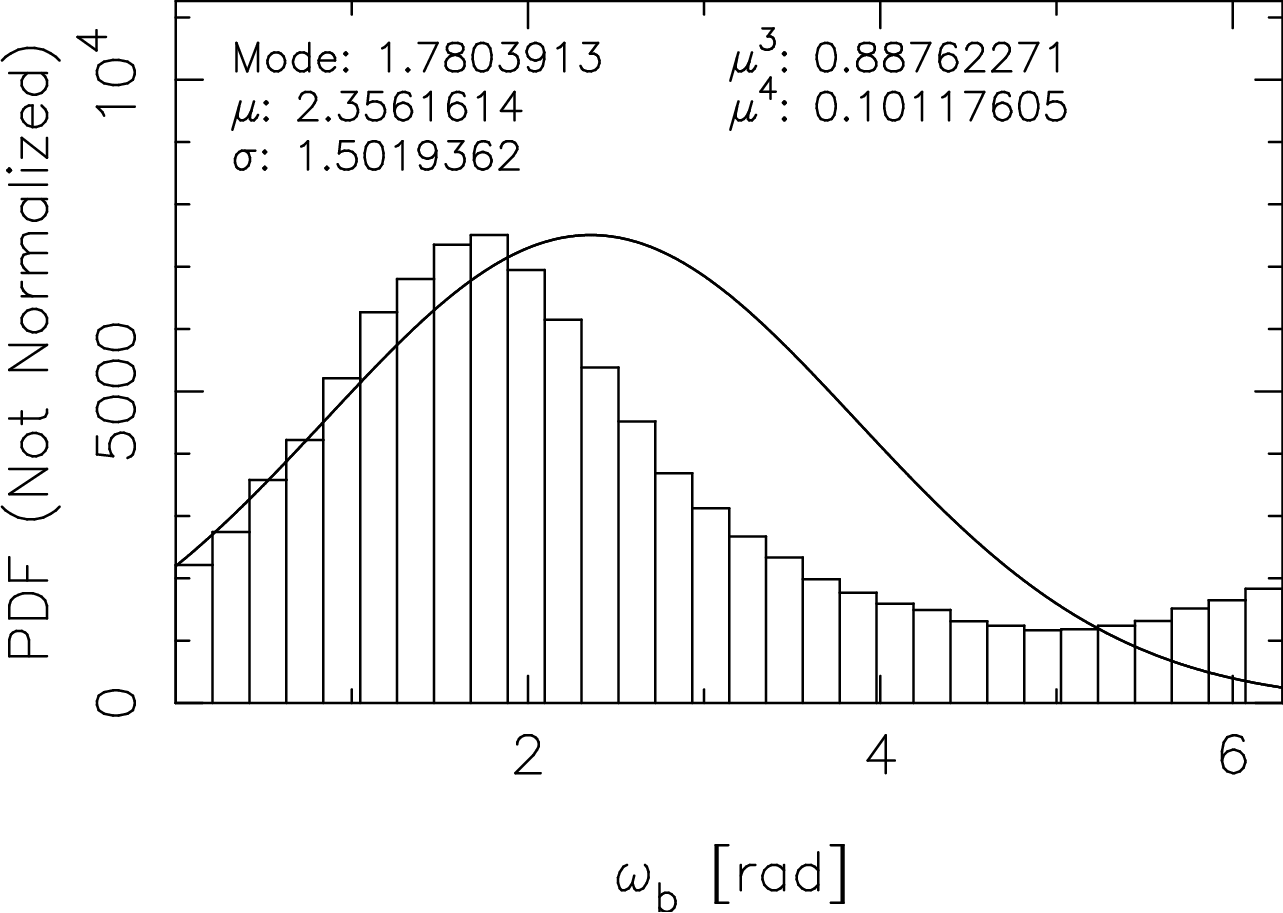}
\includegraphics[angle=-90,width=0.19\textwidth]{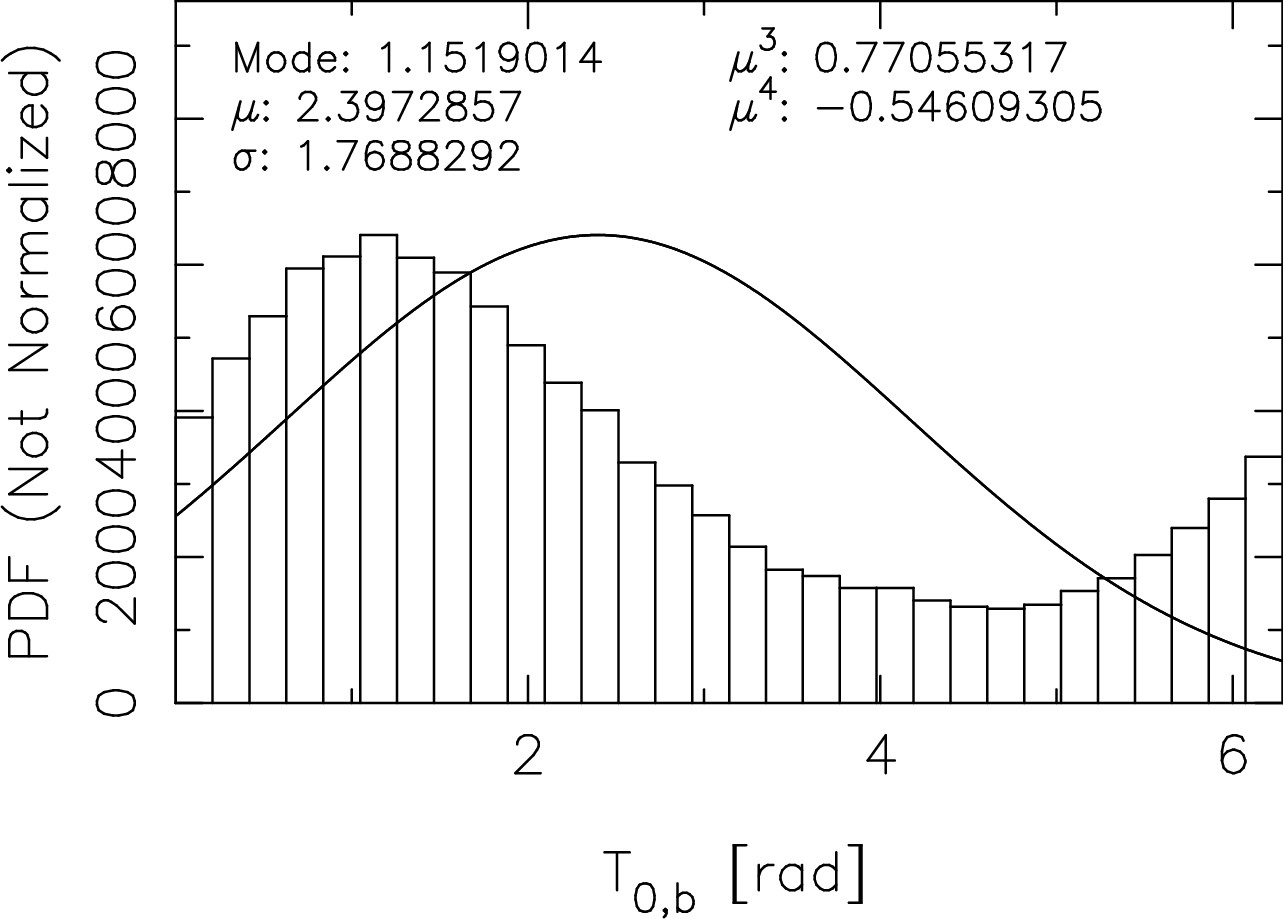}

\includegraphics[angle=-90,width=0.19\textwidth]{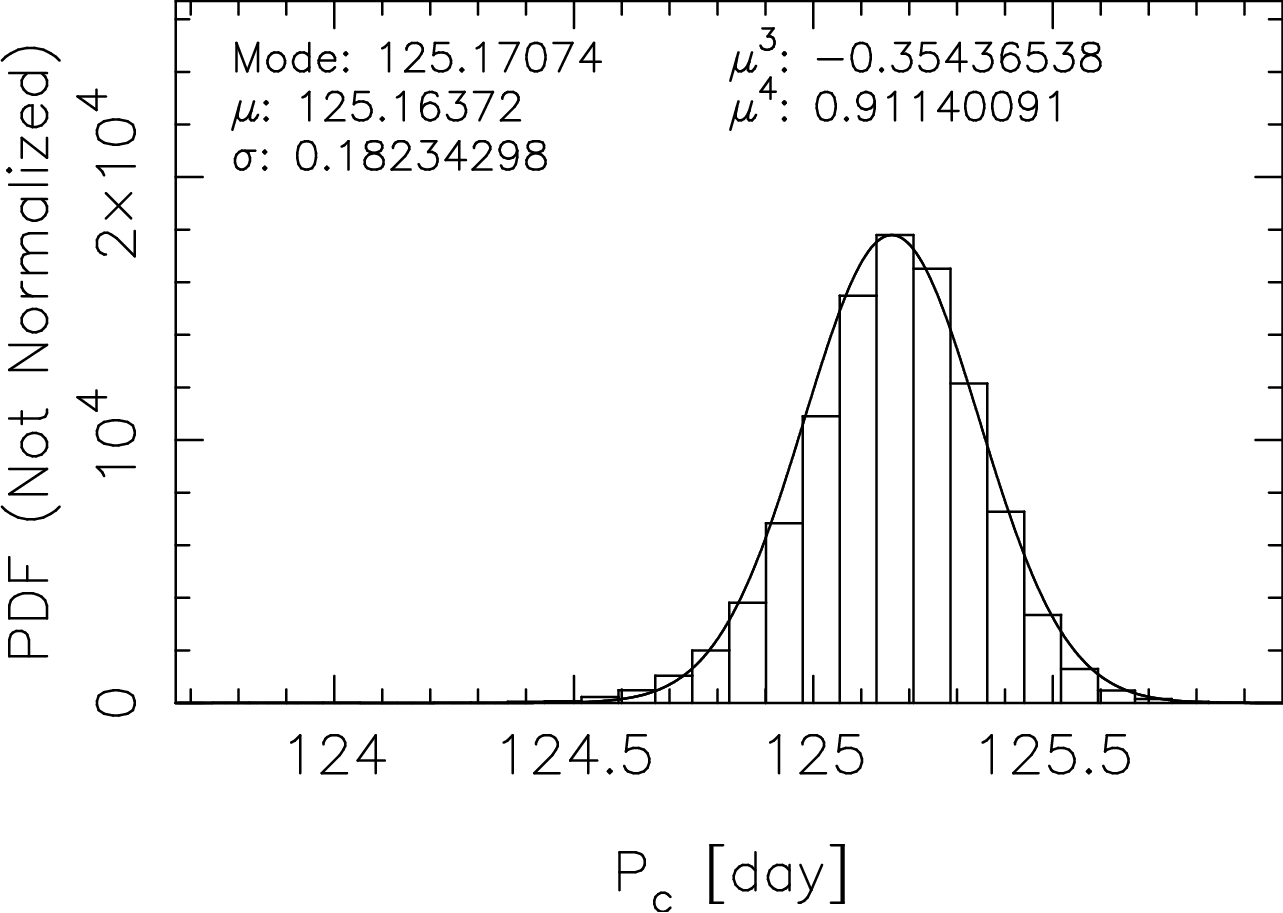}
\includegraphics[angle=-90,width=0.19\textwidth]{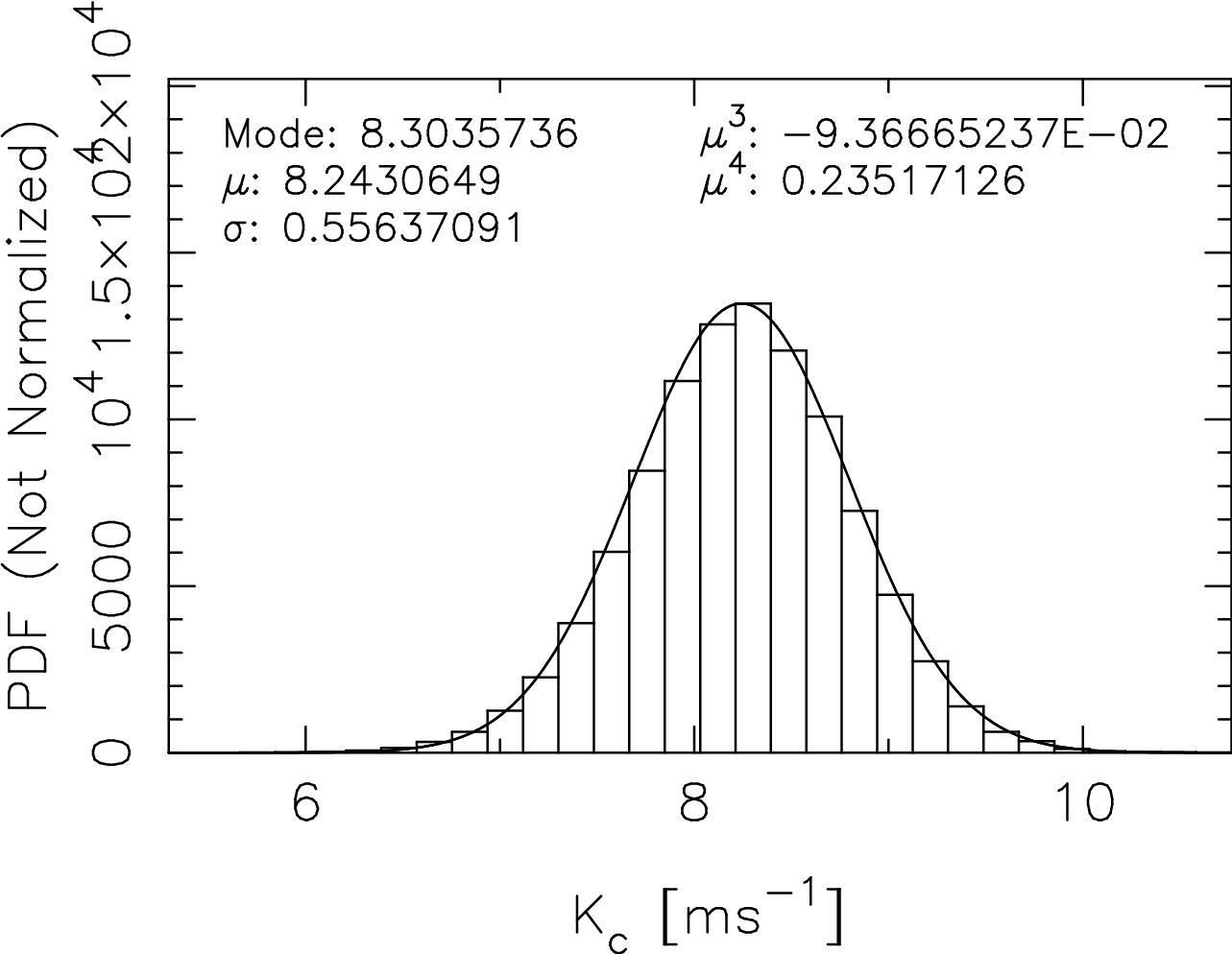}
\includegraphics[angle=-90,width=0.19\textwidth]{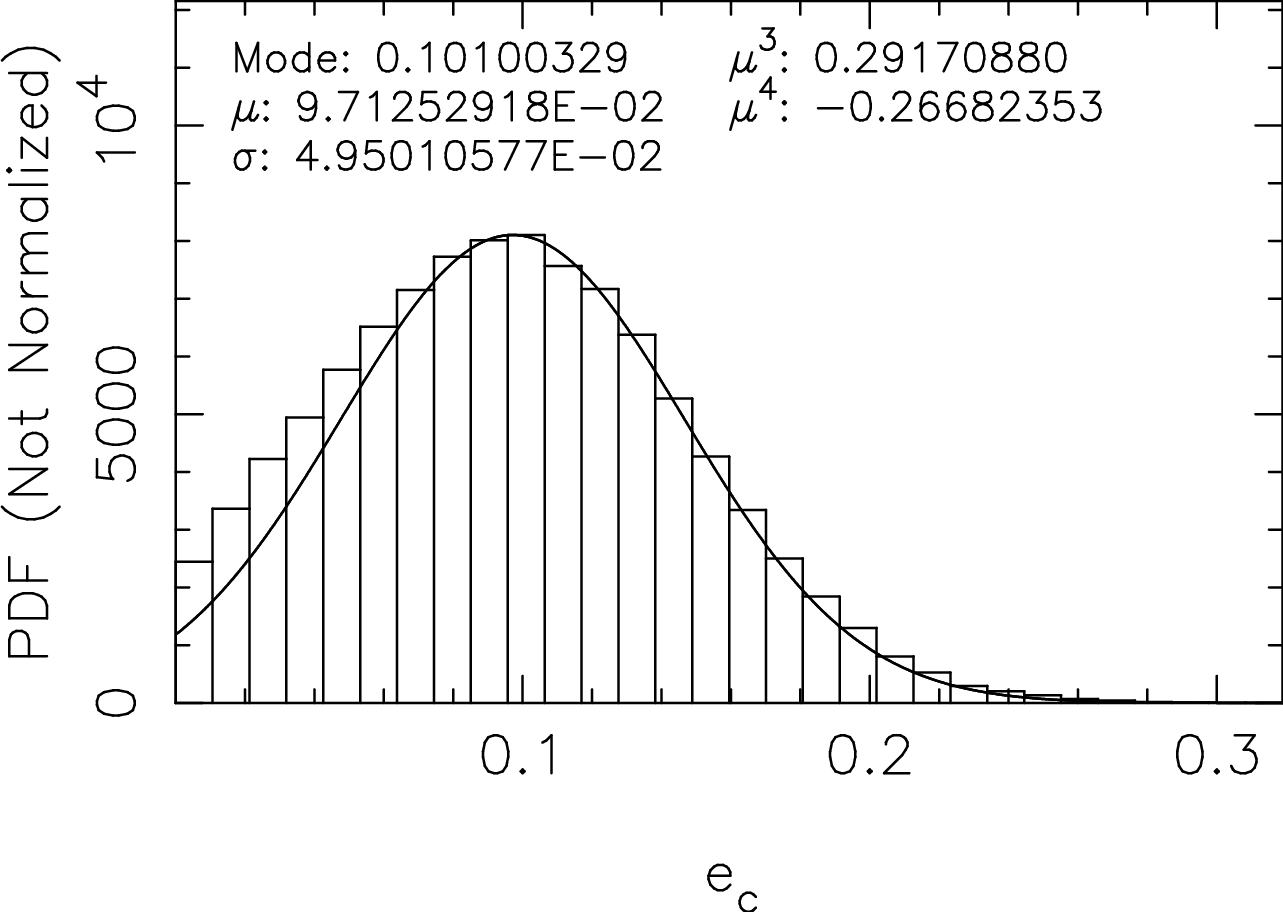}
\includegraphics[angle=-90,width=0.19\textwidth]{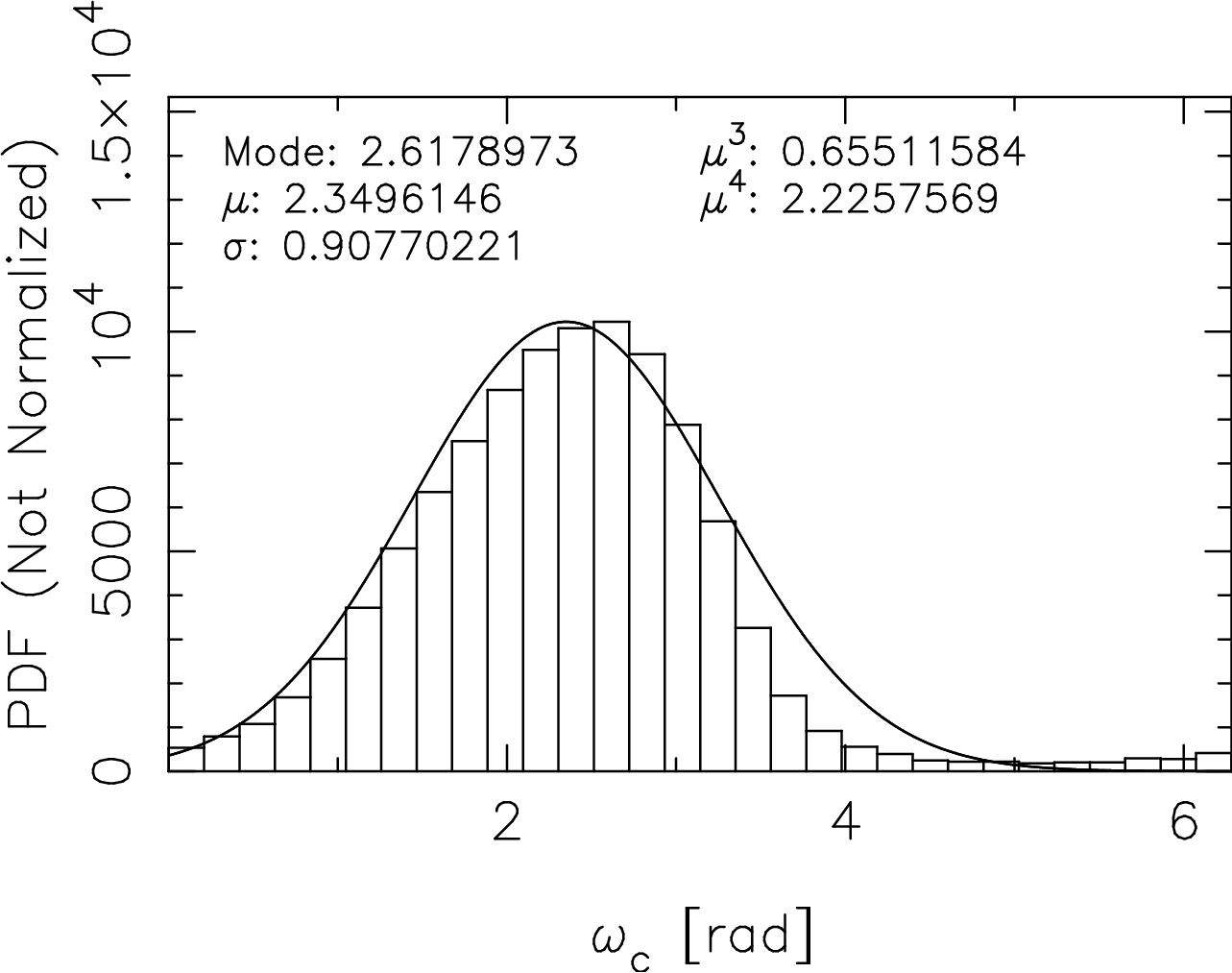}
\includegraphics[angle=-90,width=0.19\textwidth]{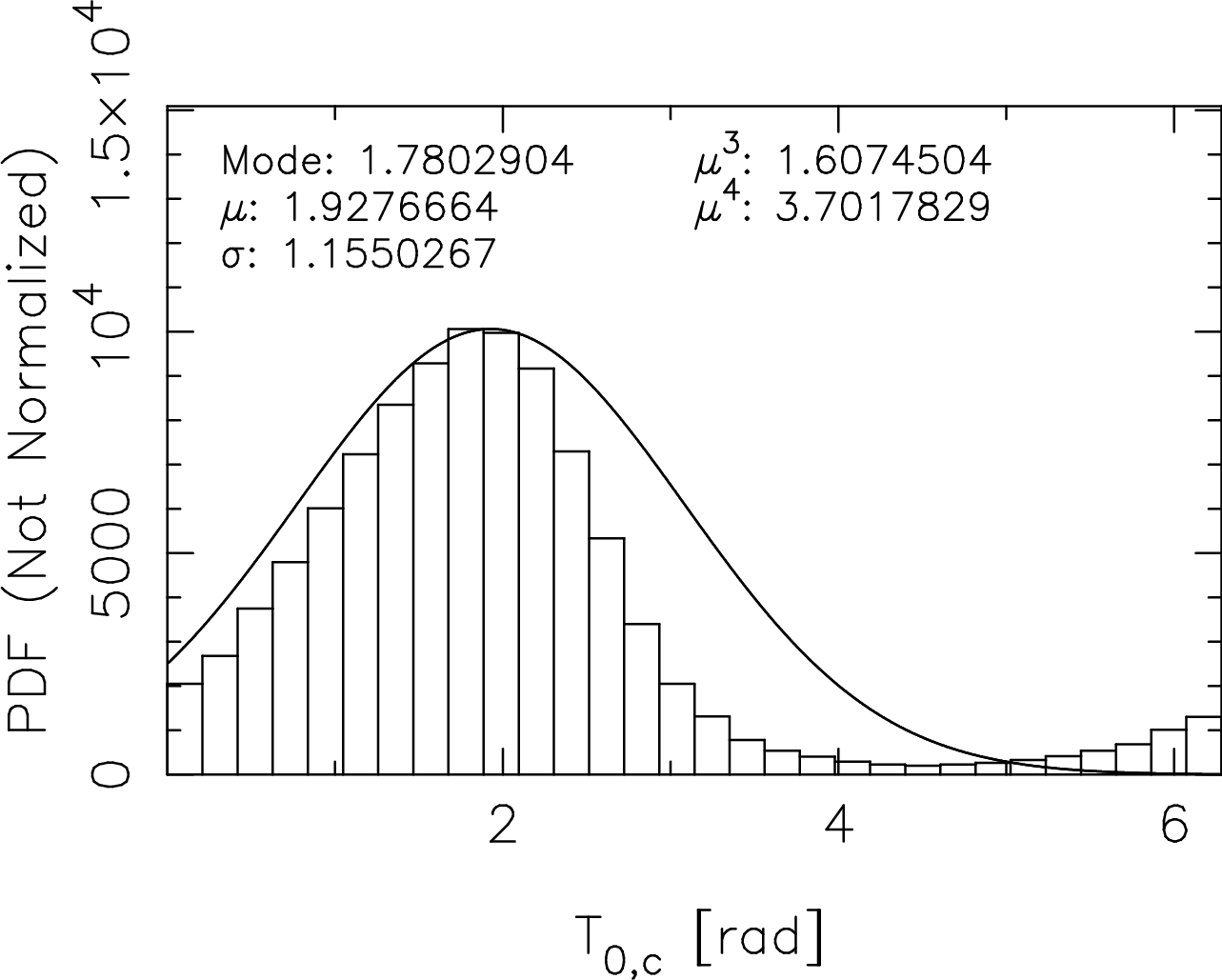}

\includegraphics[angle=-90,width=0.19\textwidth]{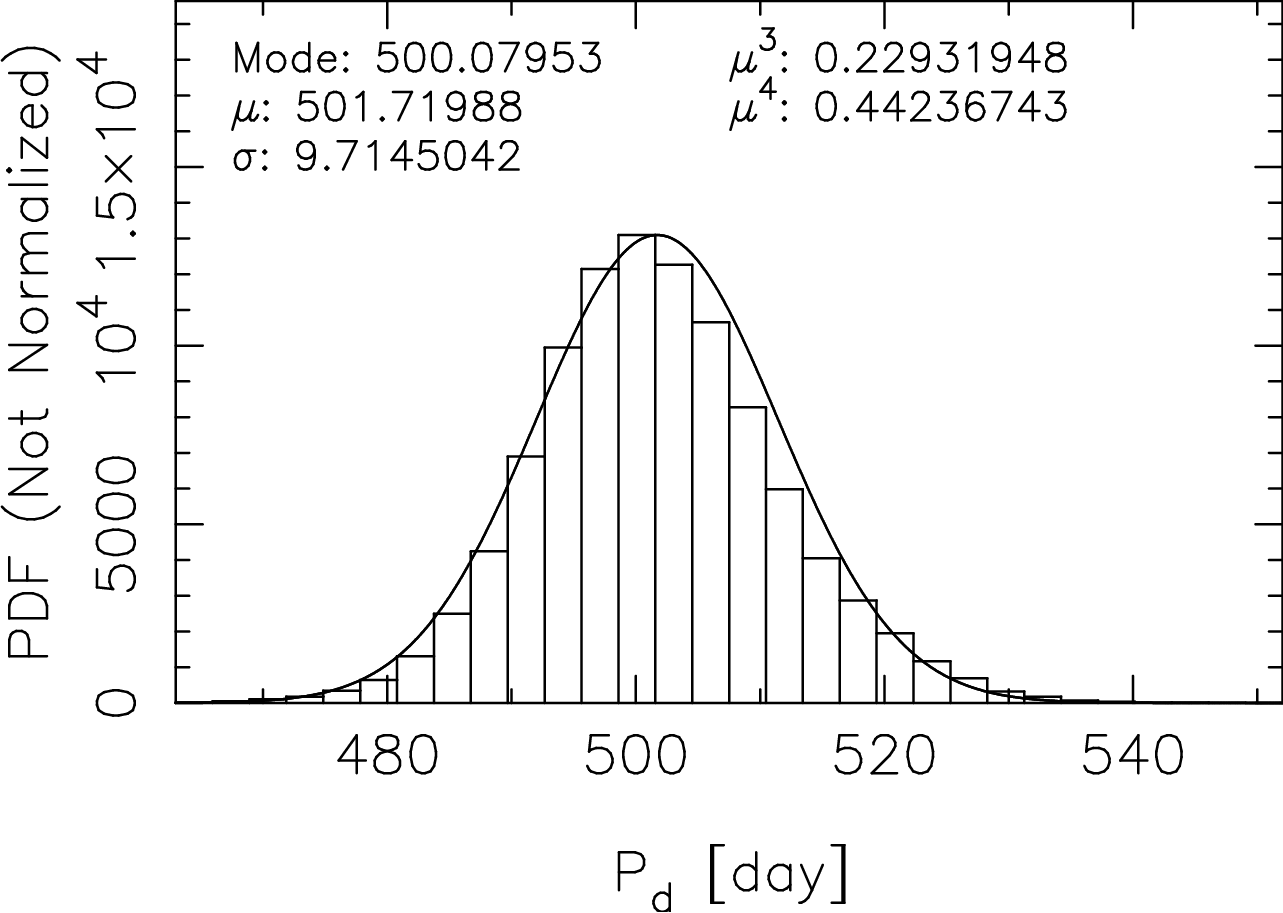}
\includegraphics[angle=-90,width=0.19\textwidth]{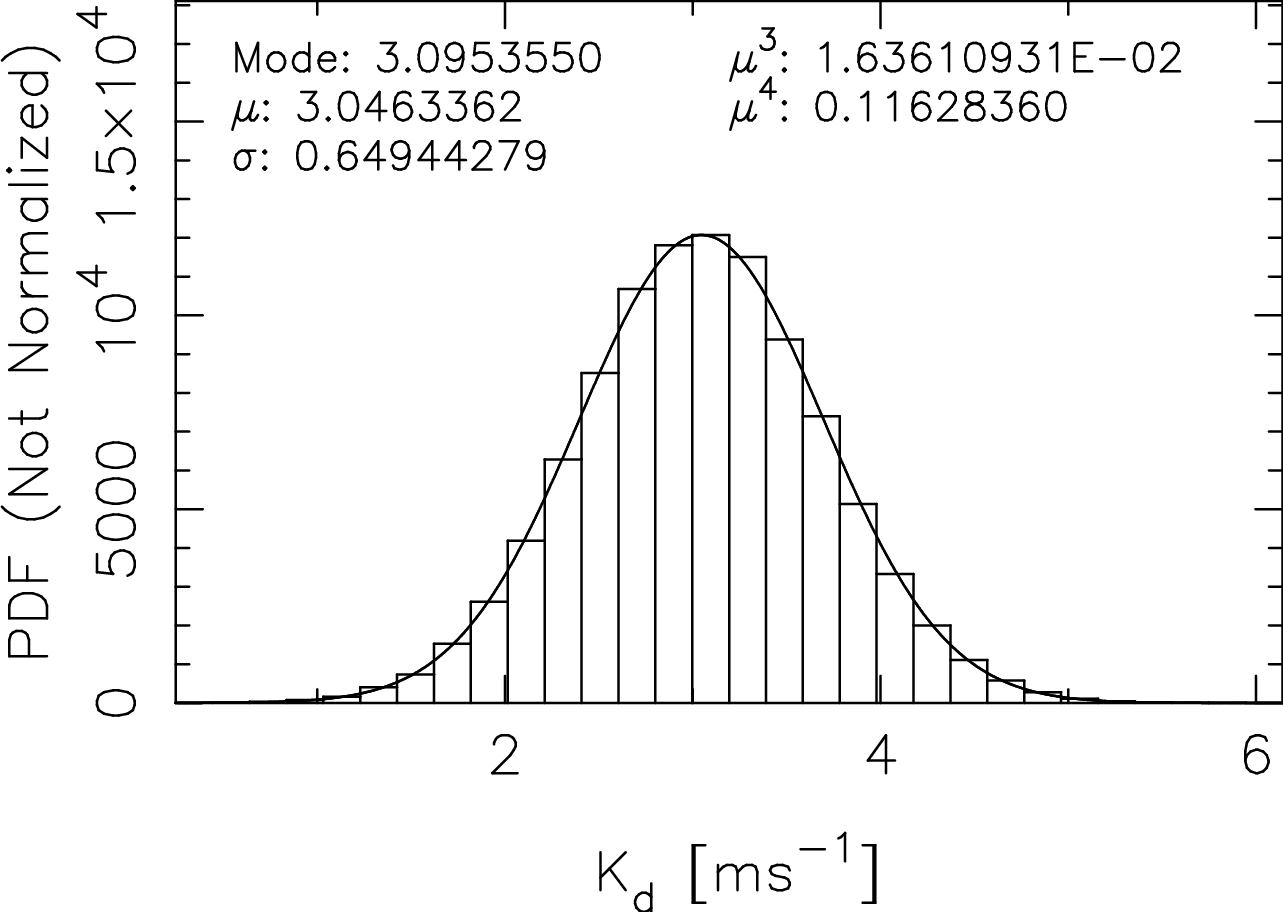}
\includegraphics[angle=-90,width=0.19\textwidth]{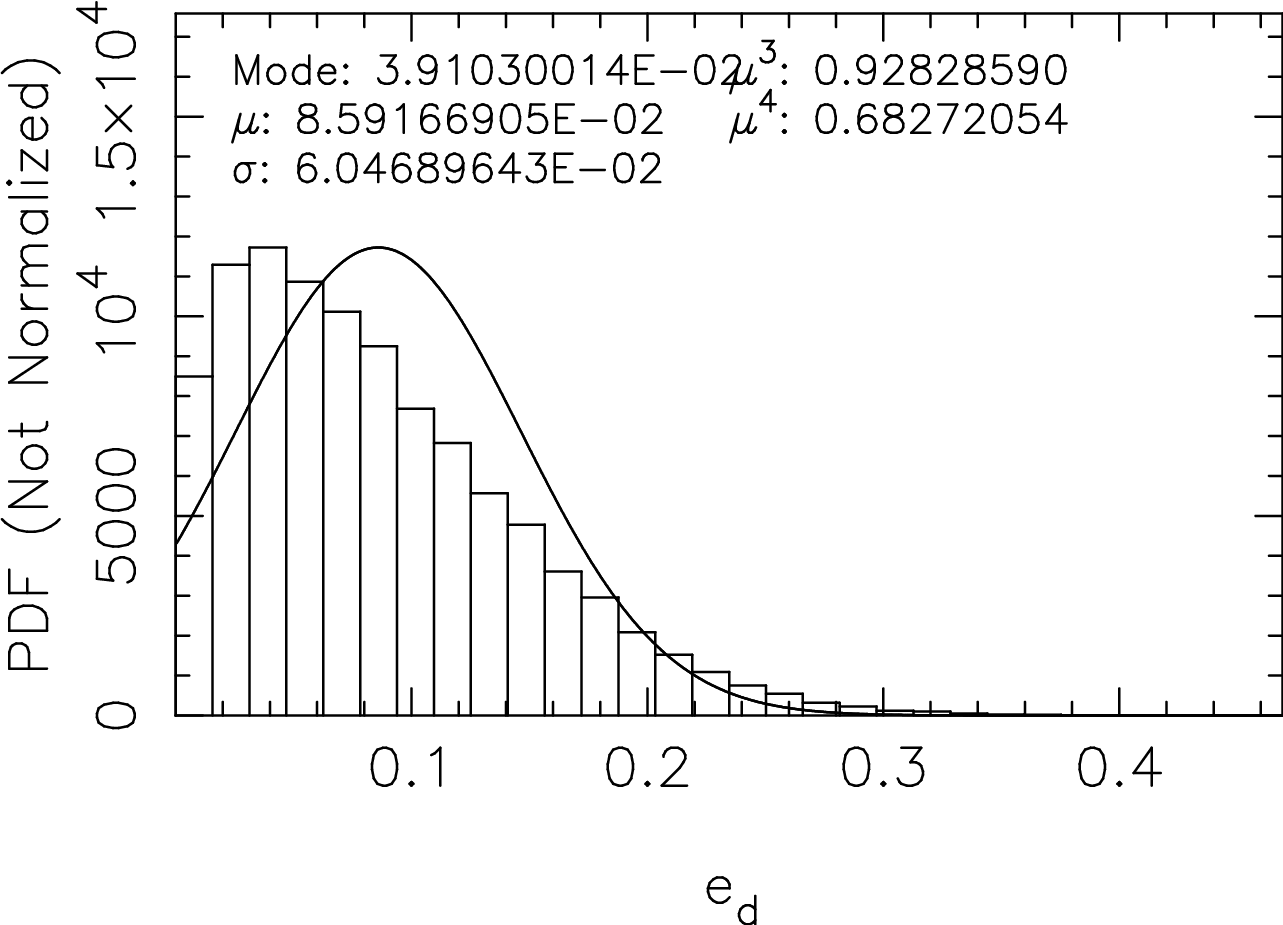}
\includegraphics[angle=-90,width=0.19\textwidth]{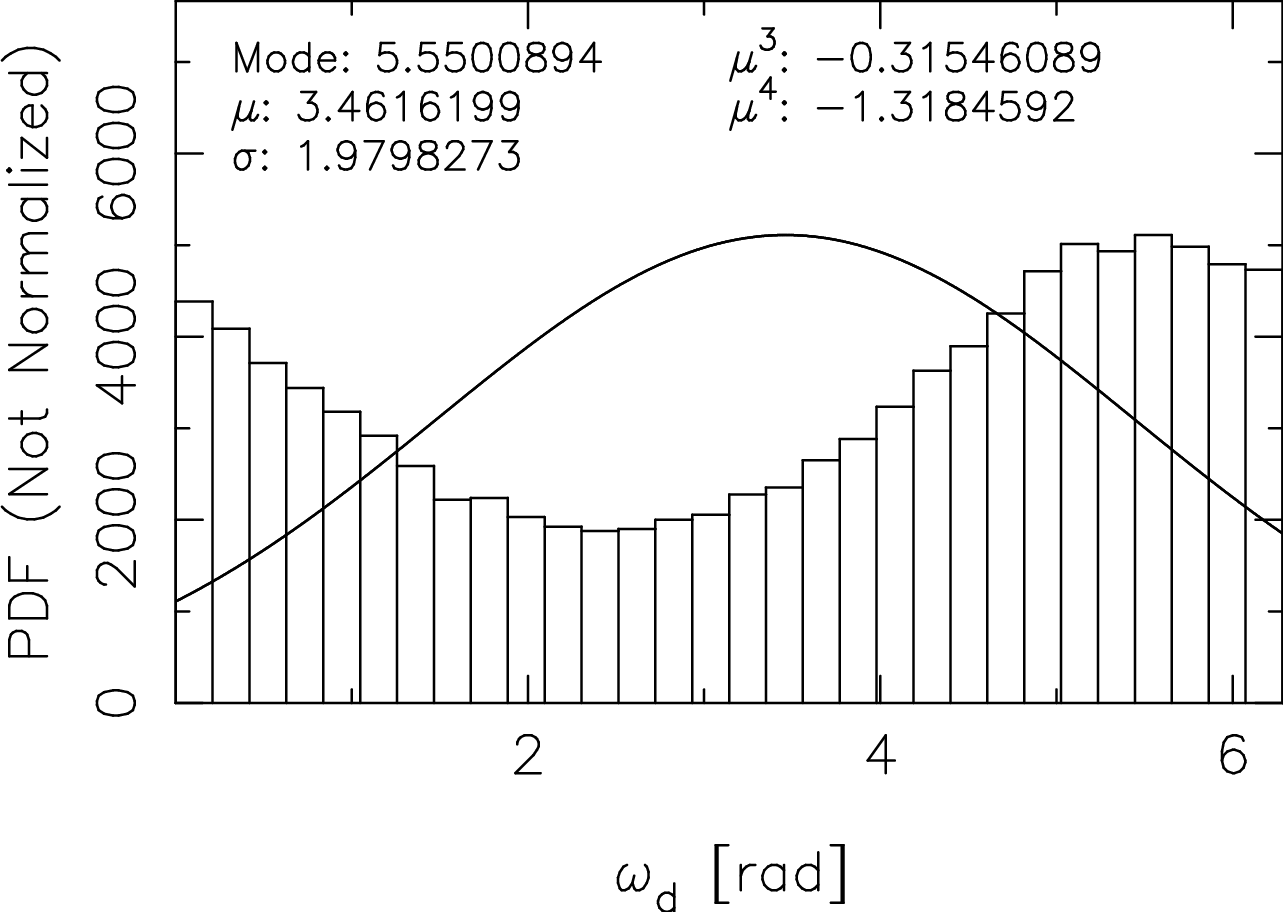}
\includegraphics[angle=-90,width=0.19\textwidth]{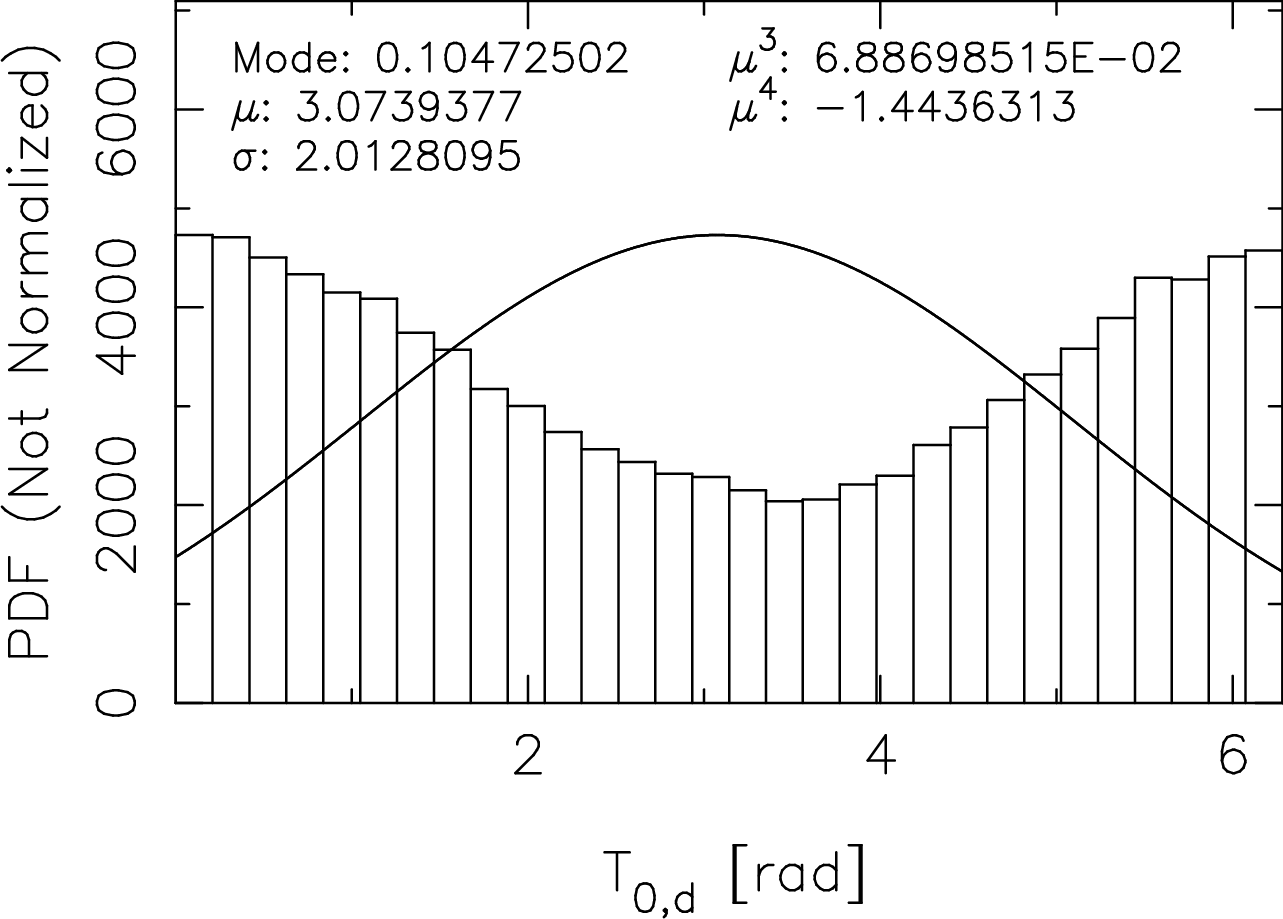}
\caption{Probability densities of the planetary orbital parameters (from left to right: period, semi-amplitude, eccentricity, longitude of the pericentre, and mean anomaly) given the HARPS-TERRA and PFS velocities. Candidates b (top row), c (middle row), and d (bottom row). The solid curves represent Gaussian densities with the same mean and variance. First four moments and the mode are shown for each density.}\label{fig:TERRA_densities}
\end{figure*}

\begin{figure}
\center
\includegraphics[angle=-90,width=0.45\textwidth]{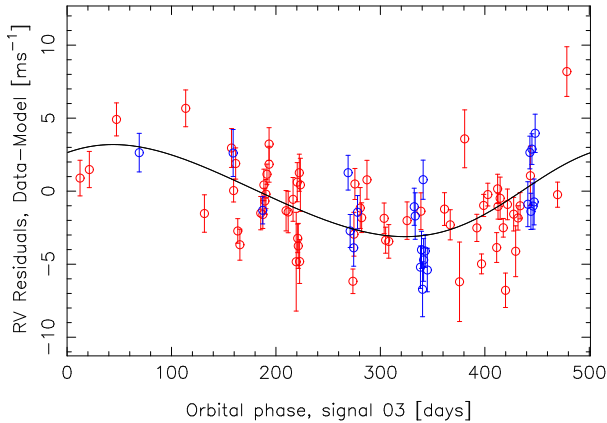}
\caption{Phase-folded signal of GJ 221 d given the combined HARPS-TERRA and PFS velocities.}\label{fig:TERRA_signal}
\end{figure}

We could consistently identify the same three periodic signals by using the combined HARPS-TERRA and PFS velocities. We show this solution in the bottom half of Table \ref{tab:parameters} and demonstrate the significance of the signals in terms of log-Bayesian evidence ratios in Table \ref{tab:probabilities}. We also plotted the posterior densities of the planetary orbital parameters in Fig. \ref{fig:TERRA_densities} and the phase-folded signal in Fig. \ref{fig:TERRA_signal}. In fact, the third signal we refer to as planet candidate GJ 221 d with an orbital period of roughly 500 days is detected slightly more condidently by using the HARPS-TERRA velocities because this third signal increases the model probability by a factor of 3.5$\times 10^{4}$ -- more than the factor of 2.6$\times 10^{4}$ that we obtained for the HARPS-CCF velocities. This small increase in significance can also be seen in Fig. \ref{fig:log-posteriors} (bottom panels) because the corresponding probability maximum is slightly more isolated one in the period space for the combined HARPS-TERRA and PFS velocities in the sense that there are no local maxima exceeding the 0.1\% probability threshold (solid lines in Fig. \ref{fig:log-posteriors}).

The estimates of the ``nuisance parameters'' of HARPS-TERRA velocities were different from those of HARPS-CCF velocities. We estimated the correlation in the HARPS-TERRA velocities to be $\phi =$ 0.24 [-0.80, 1], which is not significantly different from zero at any reasonable credibility level. This suggests that the excess noise in the TERRA velocities is likely ``whiter'' than that in the CCF data for GJ 221. The excess white noise in the TERRA data was found to be 1.92 [1.44, 3.16] ms$^{-1}$, which is consistent with the estimate for the HARPS-CCF data. Unlike for the combined PFS and HARPS-CCF data, the linear trend was significant with 95\% credibility level but not with 99\% credibility level although the estimate for the combined data with HARPS-TERRA velocities of -0.74 [-1.54, 0.04] ms$^{-1}$year$^{-1}$ (with a 99\% credibility interval) is consistent with the estimate obtained with the HARPS-CCF data. We did not find any indication of correlation in the PFS velocities.

Finally, we note that we tested a uniform eccentricity prior as well in the analyses. This did not affect the results significantly. In practice, the only effect was that the eccentricity of the candidate d was found to have a broader distribution with a MAP estimate of 0.22 and a 99\% BCS of [0, 0.69], although the eccentricity was lower than 0.4 with a 95\% probability.

\section{Discussion}

We have shown that while \citet{locurto2013} and \citet{arriagada2013} could only detect two of the planet candidates orbiting GJ 221 with the periodogram-based and related signal detection tools, with the more general techniques based on posterior samplings we were able to confidently identify the signatures of three planet candidates orbiting the star. As also demonstrated by \citet{tuomi2013c}, who detected the three candidate planets orbiting GJ 163 with $\sim$ 35\% of the data that \citet{bonfils2013b} had when announcing the discovery independently, posterior samplings provide a considerably more sensitive method for the detection of low-amplitude signals of planetary origin. This suggests that the archived data sets of various spectrographs might contain several detectable signals that have gone unnoticed due to applications of sub-optimal signal detection methods. In general, this means that the occurrence rates of low-mass planets based on radial velocity surveys might be underestimated.

We have also demonstrated that the excess noise in the HARPS-TERRA velocities of GJ 221 appears to be ``whiter'' than in the HARPS-CCF velocities in terms of having less intrinsic correlation according to the model of \citet{tuomi2013b}. Although this might also be caused by the differences in data sampling of the spectra that were not available for \citet{arriagada2013} for the computation of HARPS-TERRA veolcities, this suggests that the intrinsically better precision of the TERRA velocities might arise from this reduction in the correlated noise.

The third and previously unknown planet candidate orbiting GJ 221 is likely a cold sub-Saturnian planet with a minimum mass of 29 [9, 47] M$_{\oplus}$ and an orbital period of 500 [474, 532] days. Together with the other two planets -- the hot super-Earth with an orbital period of 3.8730 [3.8721, 3.8740] days and the warm sub-Saturnian planet with an orbital period of 125.2 [124.5, 125.7] days, GJ 221 is yet another example of the diverse population of K and M dwarfs with slightly sub-Solar metallicities that are hosts to diverse systems of multiple planetary companions. Furthermore, the configuration of three candidate planets orbiting GJ 221 is very likely a stable one because the orbital spacing satisfies the analytical Lagrange stability condition of \citet{barnes2006}. While this criterion does not guarantee stability because it is a rough approximation and does not take into account orbital resonances that might play a role in the GJ 221 system -- the orbits of the two outer planets might be affected by a 4:1 mean motion resonance -- it strongly suggests that the solution we have obtained correspond to a stable system.

Given that we could obtain almost exactly the same solution based on both HARPS-CCF and HARPS-TERRA velocities (see Table \ref{tab:parameters}), our results are independent of the exact spectral reduction method and can be considered more trustworthy than what could have been obtained based on either reduction of HARPS spectra alone.

\section*{Acknowledgements}

The author acknowledges Lo Curto et al. and Arriagada et al. for making the HARPS and PFS data public.

% ----------------------------------------------------------------

\label{lastpage}

\end{document}